\documentclass[11pt,a4paper]{article}

\renewcommand{\baselinestretch}{1.3} 

\begin{document}
\pagenumbering{roman} \setcounter{page}{0} 
\title{Scaling of Self-Avoiding Walks and Self-Avoiding Trails in Three Dimensions}
\author{T. Prellberg\thanks{{\tt {\rm email:}
thomas.prellberg@tu-clausthal.de}} \\
Institut f\"ur Theoretische Physik,\\
Technische Universit\"at Clausthal,\\ Arnold Sommerfeld Stra\ss e 6,\\ D-38678 Clausthal-Zellerfeld,\\ Germany
}
\date{\today
}

\maketitle 
 
\begin{abstract} 

Motivated by recent claims of a proof that the length scale exponent for 
the end-to-end distance scaling of self-avoiding walks is precisely 
$7/12=0.5833\ldots$, 
we present results of large-scale simulations of self-avoiding walks 
and self-avoiding trails with repulsive contact interactions
on the hypercubic lattice. We find no evidence to support this claim;
our estimate $\nu=0.5874(2)$ is in accord with the best previous
results from simulations.

\vspace{1cm} 
 

\end{abstract} 

\vfill

\newpage

\pagenumbering{arabic}
\setcounter{page}{1}

The lattice model of self-avoiding walks (SAW) has long been studied
by probabilists and physicists alike. It serves as a model for long
chain polymers in physics \cite{cloizeaux1990a-a}, is related to 
critical phenomena in statistical physics \cite{pelissetto2000a-a},
and is intriguing from a mathematical point of view due to its non-Markovian 
nature \cite{madras1993a-a}.

While there have been exact results regarding its critical behavior 
in dimensions other than three, until recently there has not even 
been a reasonable conjecture for the exact value of the length scale 
exponent $\nu$ in three dimensions.

Due to its physical importance, there have been many attempts to estimate
this value by a variety of theoretical and simulational work. An recent
overview of the simulational results is given in \cite{pelissetto2000a-a}. 
The most precise value to date from simulations is $\nu=0.58758(7)$ 
\cite{belohorec1997a-a}.

Quite surprisingly, in two recent preprints \cite{hueter2001a-a, hueter2001b-a}
there has been the announcement of a proof that $\nu=7/12=0.58333\ldots$ in three 
dimensions, which obviously is at odds with the values obtained by {\it any}
recent simulation. 

Clearly, estimates from simulational data suffer from finite-size
corrections, but the role of these corrections has also received good attention
\cite{li1995a-a}.
It is well known that finite-size estimates of $\nu$ from SAW {\em decrease} as 
the walk length increases, so that it may be advantageous to consider variants
of the model in which the finite-size corrections are different. One such model, 
in which the finite-size estimates of $\nu$ 
{\em increase}, is given by self-avoiding trails (SAT). This model is fairly
well established to be in the same universality class as SAW
\cite{guttmann1985b-a,guttmann1985c-a}. In two dimensions, different corrections
to scaling related to an irrelevant scaling variable have been observed \cite{guim1997a-a}.

An interpolation between the models is possible by assigning to each trail contact a repulsive 
interaction with Boltzmann weight $\omega$. 
Thus, we can study SAW and SAT in a unified picture. Moreover, by tuning the parameter
$\omega$ we can attempt to minimize corrections to scaling and improve our exponent estimates.

We present here data of just such a simulation. We have simulated interacting 
self-avoiding trails on the simple cubic lattice using the
Pruned-Enriched Rosenbluth Method (PERM) \cite{grassberger1997a-a}
in the implementation described in \cite{owczarek2001a-:a}.
We have generated up to $10^9$ samples of length $N=1024$ and $10^8$ samples of 
length $N=16384$.  While not having completely independent samples we have 
estimated the effect of the dependence and so are able to give error 
estimates for our values.

The interacting SAT model on the simple cubic lattice is defined in the following way. The lattice has 
coordination number $6$ and we consider configurations $\varphi_N$ of trails, or 
bond-avoiding walks, of length $N$ (bonds) starting from a fixed origin. Let $m_k, k = 1,
\ldots ,3$ be the number of sites of the lattice that have been visited
$k$ times by the trail so that $\sum k\, m_k=N+1$.  The partition
function of a very general interacting trail model is
\begin{equation}
Z_N(\omega_2,\omega_3) = \sum_{\varphi_N} \omega_2^{m_2} \omega_3^{m_3}\; ,
\end{equation}   
where $\omega_k$ is the Boltzmann weight associated with $k$-visited
sites. The canonical model is one where every segment of the trail at
some contact site interacts with every other segment at that site, so
that
\begin{equation}
\omega_k=\omega^{k\choose 2}\mbox{ for } k = 2,3\; ,
\end{equation}
with $\omega \equiv \omega_2$.  This implies that in our specific case 
\begin{equation}
\omega_2 = \omega, \qquad \omega_3=\omega^3\;.
\end{equation}
Changing $\omega$ from zero to one, we can interpolate between SAW and SAT.

In our simulations we calculated two measures of the polymer's average
size.  Firstly, specifying a trail by the sequence of position vectors
${{\mathbf r}_0, {\mathbf r}_1, ..., {\mathbf r}_N}$ the average mean-square
end-to-end distance is
\begin{equation}
\langle R^2_e\rangle_N = \langle ({\mathbf r}_N- {\mathbf r}_0) \cdot ({\mathbf r}_N- {\mathbf r}_0) \rangle\; .
\end{equation} 
The mean-square distance of a site occupied by the trail to the
endpoint, $\mathbf{r}_0$, is given by
\begin{equation}
\langle R_{m}^2 \rangle_N = \frac{1}{N+1}\sum_{i=0}^{N}\langle (\mathbf{r}_i - \mathbf{r}_0)\cdot(\mathbf{r}_i - \mathbf{r}_0)
\rangle\; .
\end{equation}
Generally one expects that
\begin{equation}
R^2_{N} \sim a(\omega)\: N^{2\nu}\quad\mbox{as}\quad N\rightarrow\infty
\end{equation}
with $\omega$-dependent amplitude $a(\omega)$. 
To estimate the exponent $\nu$ we use finite-size estimators $\nu_{e,N}$ and $\nu_{m,N}$ defined as
\begin{equation}
\nu_{e,N}=\frac12\log_2{R_{e,N}\over R_{e,N/2}}\qquad\mbox{and}\qquad
\nu_{m,N}=\frac12\log_2{R_{m,N}\over R_{m,N/2}}\;.
\end{equation}

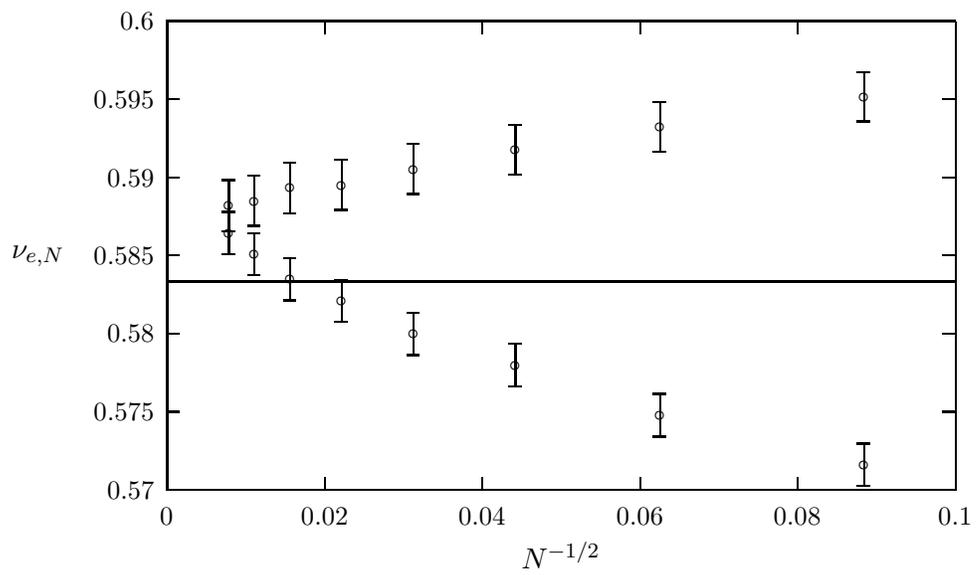
\begin{figure}
\begin{center}
\setlength{\unitlength}{0.240900pt}
\ifx\plotpoint\undefined\newsavebox{\plotpoint}\fi
\sbox{\plotpoint}{\rule[-0.200pt]{0.400pt}{0.400pt}}%
\begin{picture}(1500,900)(0,0)
\font\gnuplot=cmr10 at 10pt
\gnuplot
\sbox{\plotpoint}{\rule[-0.200pt]{0.400pt}{0.400pt}}%
\put(201.0,123.0){\rule[-0.200pt]{4.818pt}{0.400pt}}
\put(181,123){\makebox(0,0)[r]{0.57}}
\put(1419.0,123.0){\rule[-0.200pt]{4.818pt}{0.400pt}}
\put(201.0,246.0){\rule[-0.200pt]{4.818pt}{0.400pt}}
\put(181,246){\makebox(0,0)[r]{0.575}}
\put(1419.0,246.0){\rule[-0.200pt]{4.818pt}{0.400pt}}
\put(201.0,369.0){\rule[-0.200pt]{4.818pt}{0.400pt}}
\put(181,369){\makebox(0,0)[r]{0.58}}
\put(1419.0,369.0){\rule[-0.200pt]{4.818pt}{0.400pt}}
\put(201.0,492.0){\rule[-0.200pt]{4.818pt}{0.400pt}}
\put(181,492){\makebox(0,0)[r]{0.585}}
\put(1419.0,492.0){\rule[-0.200pt]{4.818pt}{0.400pt}}
\put(201.0,614.0){\rule[-0.200pt]{4.818pt}{0.400pt}}
\put(181,614){\makebox(0,0)[r]{0.59}}
\put(1419.0,614.0){\rule[-0.200pt]{4.818pt}{0.400pt}}
\put(201.0,737.0){\rule[-0.200pt]{4.818pt}{0.400pt}}
\put(181,737){\makebox(0,0)[r]{0.595}}
\put(1419.0,737.0){\rule[-0.200pt]{4.818pt}{0.400pt}}
\put(201.0,860.0){\rule[-0.200pt]{4.818pt}{0.400pt}}
\put(181,860){\makebox(0,0)[r]{0.6}}
\put(1419.0,860.0){\rule[-0.200pt]{4.818pt}{0.400pt}}
\put(201.0,123.0){\rule[-0.200pt]{0.400pt}{4.818pt}}
\put(201,82){\makebox(0,0){0}}
\put(201.0,840.0){\rule[-0.200pt]{0.400pt}{4.818pt}}
\put(449.0,123.0){\rule[-0.200pt]{0.400pt}{4.818pt}}
\put(449,82){\makebox(0,0){0.02}}
\put(449.0,840.0){\rule[-0.200pt]{0.400pt}{4.818pt}}
\put(696.0,123.0){\rule[-0.200pt]{0.400pt}{4.818pt}}
\put(696,82){\makebox(0,0){0.04}}
\put(696.0,840.0){\rule[-0.200pt]{0.400pt}{4.818pt}}
\put(944.0,123.0){\rule[-0.200pt]{0.400pt}{4.818pt}}
\put(944,82){\makebox(0,0){0.06}}
\put(944.0,840.0){\rule[-0.200pt]{0.400pt}{4.818pt}}
\put(1191.0,123.0){\rule[-0.200pt]{0.400pt}{4.818pt}}
\put(1191,82){\makebox(0,0){0.08}}
\put(1191.0,840.0){\rule[-0.200pt]{0.400pt}{4.818pt}}
\put(1439.0,123.0){\rule[-0.200pt]{0.400pt}{4.818pt}}
\put(1439,82){\makebox(0,0){0.1}}
\put(1439.0,840.0){\rule[-0.200pt]{0.400pt}{4.818pt}}
\put(201.0,123.0){\rule[-0.200pt]{298.234pt}{0.400pt}}
\put(1439.0,123.0){\rule[-0.200pt]{0.400pt}{177.543pt}}
\put(201.0,860.0){\rule[-0.200pt]{298.234pt}{0.400pt}}
\put(40,491){\makebox(0,0){$\nu_{e,N}\rule{7mm}{0pt}$}}
\put(820,21){\makebox(0,0){$N^{-1/2}$}}
\put(201.0,123.0){\rule[-0.200pt]{0.400pt}{177.543pt}}
\put(1295.0,702.0){\rule[-0.200pt]{0.400pt}{18.549pt}}
\put(1285.0,702.0){\rule[-0.200pt]{4.818pt}{0.400pt}}
\put(1285.0,779.0){\rule[-0.200pt]{4.818pt}{0.400pt}}
\put(975.0,655.0){\rule[-0.200pt]{0.400pt}{18.790pt}}
\put(965.0,655.0){\rule[-0.200pt]{4.818pt}{0.400pt}}
\put(965.0,733.0){\rule[-0.200pt]{4.818pt}{0.400pt}}
\put(748.0,619.0){\rule[-0.200pt]{0.400pt}{18.790pt}}
\put(738.0,619.0){\rule[-0.200pt]{4.818pt}{0.400pt}}
\put(738.0,697.0){\rule[-0.200pt]{4.818pt}{0.400pt}}
\put(588.0,588.0){\rule[-0.200pt]{0.400pt}{19.031pt}}
\put(578.0,588.0){\rule[-0.200pt]{4.818pt}{0.400pt}}
\put(578.0,667.0){\rule[-0.200pt]{4.818pt}{0.400pt}}
\put(475.0,563.0){\rule[-0.200pt]{0.400pt}{19.031pt}}
\put(465.0,563.0){\rule[-0.200pt]{4.818pt}{0.400pt}}
\put(465.0,642.0){\rule[-0.200pt]{4.818pt}{0.400pt}}
\put(394.0,558.0){\rule[-0.200pt]{0.400pt}{19.031pt}}
\put(384.0,558.0){\rule[-0.200pt]{4.818pt}{0.400pt}}
\put(384.0,637.0){\rule[-0.200pt]{4.818pt}{0.400pt}}
\put(338.0,538.0){\rule[-0.200pt]{0.400pt}{19.031pt}}
\put(328.0,538.0){\rule[-0.200pt]{4.818pt}{0.400pt}}
\put(328.0,617.0){\rule[-0.200pt]{4.818pt}{0.400pt}}
\put(298.0,530.0){\rule[-0.200pt]{0.400pt}{19.272pt}}
\put(288.0,530.0){\rule[-0.200pt]{4.818pt}{0.400pt}}
\put(1295,740){\circle{12}}
\put(975,694){\circle{12}}
\put(748,658){\circle{12}}
\put(588,627){\circle{12}}
\put(475,602){\circle{12}}
\put(394,598){\circle{12}}
\put(338,577){\circle{12}}
\put(298,570){\circle{12}}
\put(288.0,610.0){\rule[-0.200pt]{4.818pt}{0.400pt}}
\put(1295.0,129.0){\rule[-0.200pt]{0.400pt}{16.140pt}}
\put(1285.0,129.0){\rule[-0.200pt]{4.818pt}{0.400pt}}
\put(1285.0,196.0){\rule[-0.200pt]{4.818pt}{0.400pt}}
\put(975.0,207.0){\rule[-0.200pt]{0.400pt}{16.140pt}}
\put(965.0,207.0){\rule[-0.200pt]{4.818pt}{0.400pt}}
\put(965.0,274.0){\rule[-0.200pt]{4.818pt}{0.400pt}}
\put(748.0,286.0){\rule[-0.200pt]{0.400pt}{16.140pt}}
\put(738.0,286.0){\rule[-0.200pt]{4.818pt}{0.400pt}}
\put(738.0,353.0){\rule[-0.200pt]{4.818pt}{0.400pt}}
\put(588.0,335.0){\rule[-0.200pt]{0.400pt}{15.899pt}}
\put(578.0,335.0){\rule[-0.200pt]{4.818pt}{0.400pt}}
\put(578.0,401.0){\rule[-0.200pt]{4.818pt}{0.400pt}}
\put(475.0,387.0){\rule[-0.200pt]{0.400pt}{15.899pt}}
\put(465.0,387.0){\rule[-0.200pt]{4.818pt}{0.400pt}}
\put(465.0,453.0){\rule[-0.200pt]{4.818pt}{0.400pt}}
\put(394.0,421.0){\rule[-0.200pt]{0.400pt}{15.899pt}}
\put(384.0,421.0){\rule[-0.200pt]{4.818pt}{0.400pt}}
\put(384.0,487.0){\rule[-0.200pt]{4.818pt}{0.400pt}}
\put(338.0,461.0){\rule[-0.200pt]{0.400pt}{15.658pt}}
\put(328.0,461.0){\rule[-0.200pt]{4.818pt}{0.400pt}}
\put(328.0,526.0){\rule[-0.200pt]{4.818pt}{0.400pt}}
\put(298.0,494.0){\rule[-0.200pt]{0.400pt}{15.899pt}}
\put(288.0,494.0){\rule[-0.200pt]{4.818pt}{0.400pt}}
\put(1295,162){\circle{12}}
\put(975,241){\circle{12}}
\put(748,319){\circle{12}}
\put(588,368){\circle{12}}
\put(475,420){\circle{12}}
\put(394,454){\circle{12}}
\put(338,494){\circle{12}}
\put(298,527){\circle{12}}
\put(288.0,560.0){\rule[-0.200pt]{4.818pt}{0.400pt}}
\sbox{\plotpoint}{\rule[-0.400pt]{0.800pt}{0.800pt}}%
\put(201,451){\usebox{\plotpoint}}
\put(201.0,451.0){\rule[-0.400pt]{298.234pt}{0.800pt}}
\end{picture}
\end{center}
\caption{Finite size estimates $\nu_{e,N}$ of the length scale exponent for SAW (upper values) 
and SAT (lower values) along with the conjectured value $7/12$.}
\label{fig1}
\end{figure}

\begin{figure}
\begin{center}
\setlength{\unitlength}{0.240900pt}
\ifx\plotpoint\undefined\newsavebox{\plotpoint}\fi
\sbox{\plotpoint}{\rule[-0.200pt]{0.400pt}{0.400pt}}%
\begin{picture}(1500,900)(0,0)
\font\gnuplot=cmr10 at 10pt
\gnuplot
\sbox{\plotpoint}{\rule[-0.200pt]{0.400pt}{0.400pt}}%
\put(201.0,123.0){\rule[-0.200pt]{4.818pt}{0.400pt}}
\put(181,123){\makebox(0,0)[r]{0.57}}
\put(1419.0,123.0){\rule[-0.200pt]{4.818pt}{0.400pt}}
\put(201.0,246.0){\rule[-0.200pt]{4.818pt}{0.400pt}}
\put(181,246){\makebox(0,0)[r]{0.575}}
\put(1419.0,246.0){\rule[-0.200pt]{4.818pt}{0.400pt}}
\put(201.0,369.0){\rule[-0.200pt]{4.818pt}{0.400pt}}
\put(181,369){\makebox(0,0)[r]{0.58}}
\put(1419.0,369.0){\rule[-0.200pt]{4.818pt}{0.400pt}}
\put(201.0,492.0){\rule[-0.200pt]{4.818pt}{0.400pt}}
\put(181,492){\makebox(0,0)[r]{0.585}}
\put(1419.0,492.0){\rule[-0.200pt]{4.818pt}{0.400pt}}
\put(201.0,614.0){\rule[-0.200pt]{4.818pt}{0.400pt}}
\put(181,614){\makebox(0,0)[r]{0.59}}
\put(1419.0,614.0){\rule[-0.200pt]{4.818pt}{0.400pt}}
\put(201.0,737.0){\rule[-0.200pt]{4.818pt}{0.400pt}}
\put(181,737){\makebox(0,0)[r]{0.595}}
\put(1419.0,737.0){\rule[-0.200pt]{4.818pt}{0.400pt}}
\put(201.0,860.0){\rule[-0.200pt]{4.818pt}{0.400pt}}
\put(181,860){\makebox(0,0)[r]{0.6}}
\put(1419.0,860.0){\rule[-0.200pt]{4.818pt}{0.400pt}}
\put(201.0,123.0){\rule[-0.200pt]{0.400pt}{4.818pt}}
\put(201,82){\makebox(0,0){0}}
\put(201.0,840.0){\rule[-0.200pt]{0.400pt}{4.818pt}}
\put(449.0,123.0){\rule[-0.200pt]{0.400pt}{4.818pt}}
\put(449,82){\makebox(0,0){0.02}}
\put(449.0,840.0){\rule[-0.200pt]{0.400pt}{4.818pt}}
\put(696.0,123.0){\rule[-0.200pt]{0.400pt}{4.818pt}}
\put(696,82){\makebox(0,0){0.04}}
\put(696.0,840.0){\rule[-0.200pt]{0.400pt}{4.818pt}}
\put(944.0,123.0){\rule[-0.200pt]{0.400pt}{4.818pt}}
\put(944,82){\makebox(0,0){0.06}}
\put(944.0,840.0){\rule[-0.200pt]{0.400pt}{4.818pt}}
\put(1191.0,123.0){\rule[-0.200pt]{0.400pt}{4.818pt}}
\put(1191,82){\makebox(0,0){0.08}}
\put(1191.0,840.0){\rule[-0.200pt]{0.400pt}{4.818pt}}
\put(1439.0,123.0){\rule[-0.200pt]{0.400pt}{4.818pt}}
\put(1439,82){\makebox(0,0){0.1}}
\put(1439.0,840.0){\rule[-0.200pt]{0.400pt}{4.818pt}}
\put(201.0,123.0){\rule[-0.200pt]{298.234pt}{0.400pt}}
\put(1439.0,123.0){\rule[-0.200pt]{0.400pt}{177.543pt}}
\put(201.0,860.0){\rule[-0.200pt]{298.234pt}{0.400pt}}
\put(40,491){\makebox(0,0){$\nu_{e,N}\rule{7mm}{0pt}$}}
\put(820,21){\makebox(0,0){$N^{-1/2}$}}
\put(201.0,123.0){\rule[-0.200pt]{0.400pt}{177.543pt}}
\put(1295.0,776.0){\rule[-0.200pt]{0.400pt}{14.695pt}}
\put(1285.0,776.0){\rule[-0.200pt]{4.818pt}{0.400pt}}
\put(1285.0,837.0){\rule[-0.200pt]{4.818pt}{0.400pt}}
\put(975.0,709.0){\rule[-0.200pt]{0.400pt}{14.936pt}}
\put(965.0,709.0){\rule[-0.200pt]{4.818pt}{0.400pt}}
\put(965.0,771.0){\rule[-0.200pt]{4.818pt}{0.400pt}}
\put(748.0,657.0){\rule[-0.200pt]{0.400pt}{15.177pt}}
\put(738.0,657.0){\rule[-0.200pt]{4.818pt}{0.400pt}}
\put(738.0,720.0){\rule[-0.200pt]{4.818pt}{0.400pt}}
\put(588.0,620.0){\rule[-0.200pt]{0.400pt}{14.936pt}}
\put(578.0,620.0){\rule[-0.200pt]{4.818pt}{0.400pt}}
\put(578.0,682.0){\rule[-0.200pt]{4.818pt}{0.400pt}}
\put(475.0,590.0){\rule[-0.200pt]{0.400pt}{15.177pt}}
\put(465.0,590.0){\rule[-0.200pt]{4.818pt}{0.400pt}}
\put(465.0,653.0){\rule[-0.200pt]{4.818pt}{0.400pt}}
\put(394.0,575.0){\rule[-0.200pt]{0.400pt}{15.177pt}}
\put(384.0,575.0){\rule[-0.200pt]{4.818pt}{0.400pt}}
\put(384.0,638.0){\rule[-0.200pt]{4.818pt}{0.400pt}}
\put(338.0,539.0){\rule[-0.200pt]{0.400pt}{15.177pt}}
\put(328.0,539.0){\rule[-0.200pt]{4.818pt}{0.400pt}}
\put(328.0,602.0){\rule[-0.200pt]{4.818pt}{0.400pt}}
\put(298.0,546.0){\rule[-0.200pt]{0.400pt}{15.418pt}}
\put(288.0,546.0){\rule[-0.200pt]{4.818pt}{0.400pt}}
\put(1295,806){\circle{12}}
\put(975,740){\circle{12}}
\put(748,688){\circle{12}}
\put(588,651){\circle{12}}
\put(475,621){\circle{12}}
\put(394,606){\circle{12}}
\put(338,570){\circle{12}}
\put(298,578){\circle{12}}
\put(288.0,610.0){\rule[-0.200pt]{4.818pt}{0.400pt}}
\put(1295.0,135.0){\rule[-0.200pt]{0.400pt}{13.009pt}}
\put(1285.0,135.0){\rule[-0.200pt]{4.818pt}{0.400pt}}
\put(1285.0,189.0){\rule[-0.200pt]{4.818pt}{0.400pt}}
\put(975.0,199.0){\rule[-0.200pt]{0.400pt}{13.009pt}}
\put(965.0,199.0){\rule[-0.200pt]{4.818pt}{0.400pt}}
\put(965.0,253.0){\rule[-0.200pt]{4.818pt}{0.400pt}}
\put(748.0,272.0){\rule[-0.200pt]{0.400pt}{12.768pt}}
\put(738.0,272.0){\rule[-0.200pt]{4.818pt}{0.400pt}}
\put(738.0,325.0){\rule[-0.200pt]{4.818pt}{0.400pt}}
\put(588.0,327.0){\rule[-0.200pt]{0.400pt}{13.009pt}}
\put(578.0,327.0){\rule[-0.200pt]{4.818pt}{0.400pt}}
\put(578.0,381.0){\rule[-0.200pt]{4.818pt}{0.400pt}}
\put(475.0,372.0){\rule[-0.200pt]{0.400pt}{12.768pt}}
\put(465.0,372.0){\rule[-0.200pt]{4.818pt}{0.400pt}}
\put(465.0,425.0){\rule[-0.200pt]{4.818pt}{0.400pt}}
\put(394.0,415.0){\rule[-0.200pt]{0.400pt}{12.527pt}}
\put(384.0,415.0){\rule[-0.200pt]{4.818pt}{0.400pt}}
\put(384.0,467.0){\rule[-0.200pt]{4.818pt}{0.400pt}}
\put(338.0,454.0){\rule[-0.200pt]{0.400pt}{12.768pt}}
\put(328.0,454.0){\rule[-0.200pt]{4.818pt}{0.400pt}}
\put(328.0,507.0){\rule[-0.200pt]{4.818pt}{0.400pt}}
\put(298.0,490.0){\rule[-0.200pt]{0.400pt}{12.527pt}}
\put(288.0,490.0){\rule[-0.200pt]{4.818pt}{0.400pt}}
\put(1295,162){\circle{12}}
\put(975,226){\circle{12}}
\put(748,298){\circle{12}}
\put(588,354){\circle{12}}
\put(475,399){\circle{12}}
\put(394,441){\circle{12}}
\put(338,481){\circle{12}}
\put(298,516){\circle{12}}
\put(288.0,542.0){\rule[-0.200pt]{4.818pt}{0.400pt}}
\sbox{\plotpoint}{\rule[-0.400pt]{0.800pt}{0.800pt}}%
\put(201,451){\usebox{\plotpoint}}
\put(201.0,451.0){\rule[-0.400pt]{298.234pt}{0.800pt}}
\end{picture}
\end{center}
\caption{Finite size estimates $\nu_{m,N}$ of the length scale exponent for SAW (upper values) 
and SAT (lower values) along with the conjectured value $7/12$.}
\label{fig2}
\end{figure}
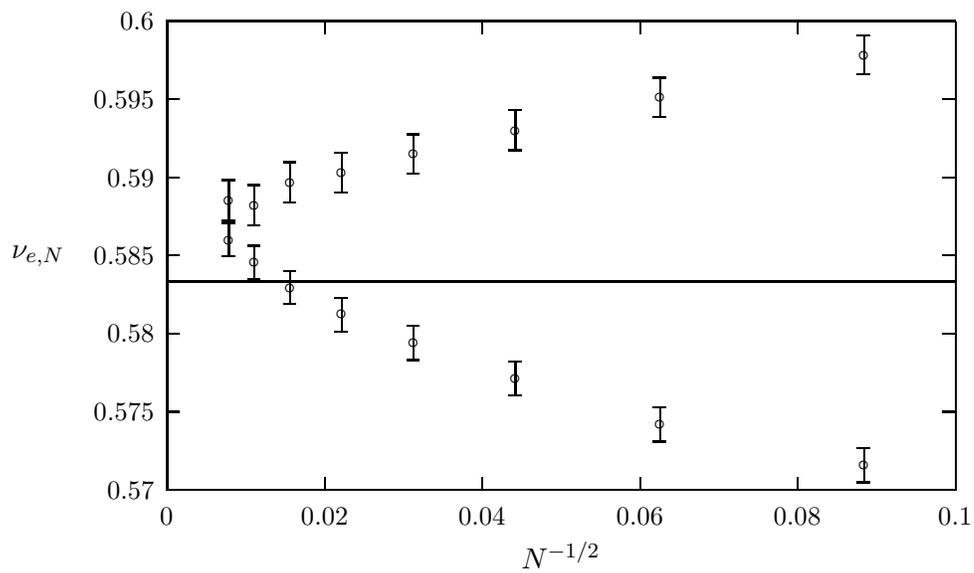

Our results for the finite-size estimates of the end-to-end distance scaling are shown in 
Figure \ref{fig1} and the for the corresponding mean-distance scaling are shown in 
Figure \ref{fig2}. We see that while finite-size estimates of $\nu$ for SAT are well
below $7/12$ for shorter trails, they cross this value around trail lengths of $N=4000$ and
are finally well above it and in correspondence with finite-size estimates of $\nu$ for SAW. 

\begin{figure}
\begin{center}
\setlength{\unitlength}{0.240900pt}
\ifx\plotpoint\undefined\newsavebox{\plotpoint}\fi
\sbox{\plotpoint}{\rule[-0.200pt]{0.400pt}{0.400pt}}%
\begin{picture}(1500,900)(0,0)
\font\gnuplot=cmr10 at 10pt
\gnuplot
\sbox{\plotpoint}{\rule[-0.200pt]{0.400pt}{0.400pt}}%
\put(201.0,123.0){\rule[-0.200pt]{4.818pt}{0.400pt}}
\put(181,123){\makebox(0,0)[r]{0.584}}
\put(1419.0,123.0){\rule[-0.200pt]{4.818pt}{0.400pt}}
\put(201.0,215.0){\rule[-0.200pt]{4.818pt}{0.400pt}}
\put(181,215){\makebox(0,0)[r]{0.585}}
\put(1419.0,215.0){\rule[-0.200pt]{4.818pt}{0.400pt}}
\put(201.0,307.0){\rule[-0.200pt]{4.818pt}{0.400pt}}
\put(181,307){\makebox(0,0)[r]{0.586}}
\put(1419.0,307.0){\rule[-0.200pt]{4.818pt}{0.400pt}}
\put(201.0,399.0){\rule[-0.200pt]{4.818pt}{0.400pt}}
\put(181,399){\makebox(0,0)[r]{0.587}}
\put(1419.0,399.0){\rule[-0.200pt]{4.818pt}{0.400pt}}
\put(201.0,491.0){\rule[-0.200pt]{4.818pt}{0.400pt}}
\put(181,491){\makebox(0,0)[r]{0.588}}
\put(1419.0,491.0){\rule[-0.200pt]{4.818pt}{0.400pt}}
\put(201.0,584.0){\rule[-0.200pt]{4.818pt}{0.400pt}}
\put(181,584){\makebox(0,0)[r]{0.589}}
\put(1419.0,584.0){\rule[-0.200pt]{4.818pt}{0.400pt}}
\put(201.0,676.0){\rule[-0.200pt]{4.818pt}{0.400pt}}
\put(181,676){\makebox(0,0)[r]{0.59}}
\put(1419.0,676.0){\rule[-0.200pt]{4.818pt}{0.400pt}}
\put(201.0,768.0){\rule[-0.200pt]{4.818pt}{0.400pt}}
\put(181,768){\makebox(0,0)[r]{0.591}}
\put(1419.0,768.0){\rule[-0.200pt]{4.818pt}{0.400pt}}
\put(201.0,860.0){\rule[-0.200pt]{4.818pt}{0.400pt}}
\put(181,860){\makebox(0,0)[r]{0.592}}
\put(1419.0,860.0){\rule[-0.200pt]{4.818pt}{0.400pt}}
\put(201.0,123.0){\rule[-0.200pt]{0.400pt}{4.818pt}}
\put(201,82){\makebox(0,0){0}}
\put(201.0,840.0){\rule[-0.200pt]{0.400pt}{4.818pt}}
\put(391.0,123.0){\rule[-0.200pt]{0.400pt}{4.818pt}}
\put(391,82){\makebox(0,0){0.02}}
\put(391.0,840.0){\rule[-0.200pt]{0.400pt}{4.818pt}}
\put(582.0,123.0){\rule[-0.200pt]{0.400pt}{4.818pt}}
\put(582,82){\makebox(0,0){0.04}}
\put(582.0,840.0){\rule[-0.200pt]{0.400pt}{4.818pt}}
\put(772.0,123.0){\rule[-0.200pt]{0.400pt}{4.818pt}}
\put(772,82){\makebox(0,0){0.06}}
\put(772.0,840.0){\rule[-0.200pt]{0.400pt}{4.818pt}}
\put(963.0,123.0){\rule[-0.200pt]{0.400pt}{4.818pt}}
\put(963,82){\makebox(0,0){0.08}}
\put(963.0,840.0){\rule[-0.200pt]{0.400pt}{4.818pt}}
\put(1153.0,123.0){\rule[-0.200pt]{0.400pt}{4.818pt}}
\put(1153,82){\makebox(0,0){0.1}}
\put(1153.0,840.0){\rule[-0.200pt]{0.400pt}{4.818pt}}
\put(1344.0,123.0){\rule[-0.200pt]{0.400pt}{4.818pt}}
\put(1344,82){\makebox(0,0){0.12}}
\put(1344.0,840.0){\rule[-0.200pt]{0.400pt}{4.818pt}}
\put(201.0,123.0){\rule[-0.200pt]{298.234pt}{0.400pt}}
\put(1439.0,123.0){\rule[-0.200pt]{0.400pt}{177.543pt}}
\put(201.0,860.0){\rule[-0.200pt]{298.234pt}{0.400pt}}
\put(40,491){\makebox(0,0){$\nu_{e,N}\rule{7mm}{0pt}$}}
\put(820,21){\makebox(0,0){$N^{-1/2}$}}
\put(201.0,123.0){\rule[-0.200pt]{0.400pt}{177.543pt}}
\put(1391.0,735.0){\rule[-0.200pt]{0.400pt}{25.776pt}}
\put(1381.0,735.0){\rule[-0.200pt]{4.818pt}{0.400pt}}
\put(1381.0,842.0){\rule[-0.200pt]{4.818pt}{0.400pt}}
\put(1043.0,665.0){\rule[-0.200pt]{0.400pt}{26.499pt}}
\put(1033.0,665.0){\rule[-0.200pt]{4.818pt}{0.400pt}}
\put(1033.0,775.0){\rule[-0.200pt]{4.818pt}{0.400pt}}
\put(796.0,598.0){\rule[-0.200pt]{0.400pt}{25.294pt}}
\put(786.0,598.0){\rule[-0.200pt]{4.818pt}{0.400pt}}
\put(786.0,703.0){\rule[-0.200pt]{4.818pt}{0.400pt}}
\put(622.0,547.0){\rule[-0.200pt]{0.400pt}{26.017pt}}
\put(612.0,547.0){\rule[-0.200pt]{4.818pt}{0.400pt}}
\put(612.0,655.0){\rule[-0.200pt]{4.818pt}{0.400pt}}
\put(499.0,527.0){\rule[-0.200pt]{0.400pt}{27.222pt}}
\put(489.0,527.0){\rule[-0.200pt]{4.818pt}{0.400pt}}
\put(1391,789){\circle{12}}
\put(1043,720){\circle{12}}
\put(796,651){\circle{12}}
\put(622,601){\circle{12}}
\put(499,584){\circle{12}}
\put(489.0,640.0){\rule[-0.200pt]{4.818pt}{0.400pt}}
\put(1391.0,419.0){\rule[-0.200pt]{0.400pt}{4.577pt}}
\put(1381.0,419.0){\rule[-0.200pt]{4.818pt}{0.400pt}}
\put(1381.0,438.0){\rule[-0.200pt]{4.818pt}{0.400pt}}
\put(1043.0,422.0){\rule[-0.200pt]{0.400pt}{4.818pt}}
\put(1033.0,422.0){\rule[-0.200pt]{4.818pt}{0.400pt}}
\put(1033.0,442.0){\rule[-0.200pt]{4.818pt}{0.400pt}}
\put(796.0,420.0){\rule[-0.200pt]{0.400pt}{5.059pt}}
\put(786.0,420.0){\rule[-0.200pt]{4.818pt}{0.400pt}}
\put(786.0,441.0){\rule[-0.200pt]{4.818pt}{0.400pt}}
\put(622.0,425.0){\rule[-0.200pt]{0.400pt}{5.059pt}}
\put(612.0,425.0){\rule[-0.200pt]{4.818pt}{0.400pt}}
\put(612.0,446.0){\rule[-0.200pt]{4.818pt}{0.400pt}}
\put(499.0,429.0){\rule[-0.200pt]{0.400pt}{5.059pt}}
\put(489.0,429.0){\rule[-0.200pt]{4.818pt}{0.400pt}}
\put(1391,429){\circle{12}}
\put(1043,432){\circle{12}}
\put(796,431){\circle{12}}
\put(622,435){\circle{12}}
\put(499,439){\circle{12}}
\put(489.0,450.0){\rule[-0.200pt]{4.818pt}{0.400pt}}
\put(1391.0,186.0){\rule[-0.200pt]{0.400pt}{14.936pt}}
\put(1381.0,186.0){\rule[-0.200pt]{4.818pt}{0.400pt}}
\put(1381.0,248.0){\rule[-0.200pt]{4.818pt}{0.400pt}}
\put(1043.0,237.0){\rule[-0.200pt]{0.400pt}{14.695pt}}
\put(1033.0,237.0){\rule[-0.200pt]{4.818pt}{0.400pt}}
\put(1033.0,298.0){\rule[-0.200pt]{4.818pt}{0.400pt}}
\put(796.0,265.0){\rule[-0.200pt]{0.400pt}{14.936pt}}
\put(786.0,265.0){\rule[-0.200pt]{4.818pt}{0.400pt}}
\put(786.0,327.0){\rule[-0.200pt]{4.818pt}{0.400pt}}
\put(622.0,303.0){\rule[-0.200pt]{0.400pt}{15.658pt}}
\put(612.0,303.0){\rule[-0.200pt]{4.818pt}{0.400pt}}
\put(612.0,368.0){\rule[-0.200pt]{4.818pt}{0.400pt}}
\put(499.0,346.0){\rule[-0.200pt]{0.400pt}{15.658pt}}
\put(489.0,346.0){\rule[-0.200pt]{4.818pt}{0.400pt}}
\put(1391,217){\circle{12}}
\put(1043,267){\circle{12}}
\put(796,296){\circle{12}}
\put(622,335){\circle{12}}
\put(499,378){\circle{12}}
\put(489.0,411.0){\rule[-0.200pt]{4.818pt}{0.400pt}}
\put(201,418){\usebox{\plotpoint}}
\put(201.0,418.0){\rule[-0.200pt]{298.234pt}{0.400pt}}
\put(201,455){\usebox{\plotpoint}}
\put(201.0,455.0){\rule[-0.200pt]{298.234pt}{0.400pt}}
\end{picture}
\end{center}
\caption{Finite size estimates $\nu_{m,N}$ of the length scale exponent for interacting SAT at
$\omega=0.40$, $\omega=0.53$, and $\omega=0.60$ from top to bottom. The horizontal lines indicate
our estimate $\nu=0.5874(2)$.
}
\label{fig3}
\end{figure}
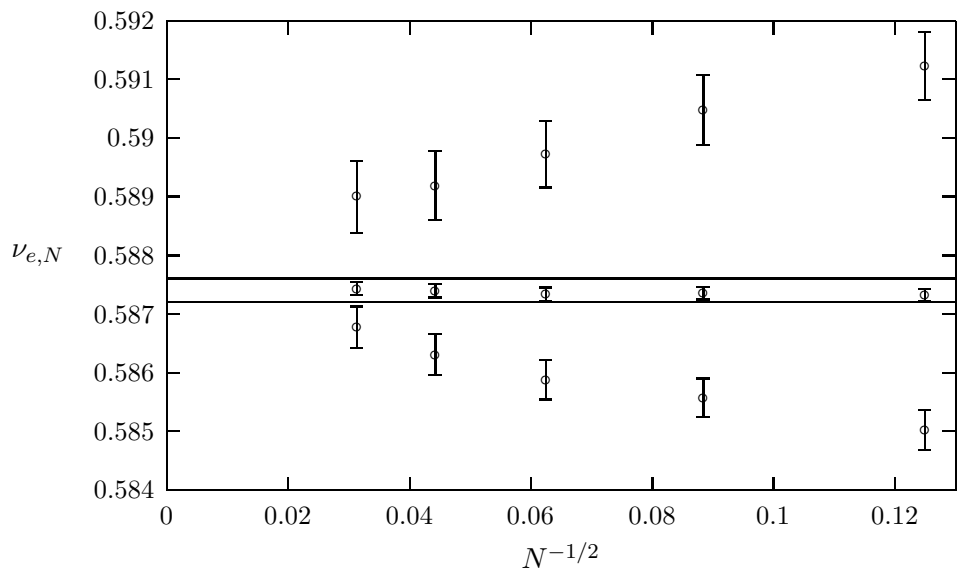

Given that the finite-size corrections have different sign for SAW and SAT, we have tried to
obtain a value of $\omega$ where the leading correction approximately vanishes in order to
obtain a better estimate from high-precision runs at shorter length. This happens around different
values of $\omega$ for $\nu_{e,N}$ and $\nu_{m,N}$. As the error bars for $\nu_{m,N}$ are
slightly smaller, we have focussed on the latter.  
For $\omega=0.53$ we find indeed that the estimators $\nu_{m,N}$ are virtually non-changing, as
shown in Figure \ref{fig3}, where also results for $\omega=0.4$ and $\omega=0.6$ is shown for comparison. 
We estimate from this that $\nu=0.5874(2)$.

In summary, we have investigated scaling properties of interacting
SAT in the repulsive regime, provided evidence that SAW and
SAT are indeed in the same universality class, and obtained the
estimate $\nu=0.5874(2)$. We find no indication of the conjectured value $7/12$.

\clearpage
\newpage

\clearpage
\newpage

\clearpage
\newpage

\end{document}